\documentclass{emulateapj}

\usepackage{epsfig}
\usepackage{lscape}

\newcommand{\etal}{\mbox{et al.}}
\newcommand{\ergcms}{erg cm$^{-2}$ s$^{-1}$}
\newcommand{\ergs}{erg s$^{-1}$}

\newcommand{\degree}{$^\circ$}
\newcommand{\msun}{$M_{\odot}$}

\newcommand{\chandra}{{\it Chandra}}

\newcommand{\asca}{{\it ASCA}}

\newcommand{\xmm}{{\it XMM-Newton}}

\newcommand{\rxte}{{\it RXTE}}

\newcommand{\gogus}{G\"{o}\u{g}\"{u}\c{s}}

\shortauthors{Muno \etal}
\shorttitle{A Search for Magnetars}

\begin{document}

\title {A Search for New Galactic Magnetars in Archival Chandra and XMM-Newton Observations}

\author{M. P. Muno,\altaffilmark{1} B. M. Gaensler,\altaffilmark{2} 
A. Nechita,\altaffilmark{3} J. M. Miller,\altaffilmark{4} and 
P. O. Slane\altaffilmark{5}} 

\altaffiltext{1}{Space Radiation Laboratory, California Institute of 
Technology, Pasadena CA 90025; mmuno@srl.caltech.edu}
\altaffiltext{2}{School of Physics A29, The University of Sydney, 
NSW 2006, Australia}
\altaffiltext{3}{536 Broad St., The Onion, New York, NY 10012}
\altaffiltext{4}{Department of Astronomy and Astrophysics, The University
of Michigan, 500 Church Street, Ann Arbor, Michigan, 48109}
\altaffiltext{5}{Harvard-Smithsonian Center for Astrophysics, Cambridge, 
MA 02138}

\begin{abstract} 
We present constraints on the number of Galactic
magnetars, which we have established by searching for sources with
periodic variability in 506 archival \chandra\ observations and 441
archival \xmm\ observations of the Galactic plane ($|b|$$<$5\degree).
Our search revealed four sources with periodic variability on time
scales of 200--5000 s, all of which are probably accreting white
dwarfs.  We identify 7 of 12 known Galactic magnetars, but find no new
examples with periods between 5 and 20 s. We convert this
non-detection into limits on the total number of Galactic magnetars by
computing the fraction of the young Galactic stellar population that
was included in our survey. We find that easily-detectable magnetars,
modeled after persistent anomalous X-ray pulsars (e.g., either with
X-ray luminosities $L_{\rm X}$=$3\times10^{33}$ \ergs\ [0.5--10.0 keV]
and fractional rms amplitudes $A_{\rm rms}$=70\%, or $L_{\rm
X}$=$10^{35}$ \ergs\ and $A_{\rm rms}$=12\%), could have been
identified in $\approx$5\% of the Galactic spiral arms by mass. If we
assume that there are 3 previously-known examples within our random survey,
then there are 59$^{+92}_{-32}$ in the
Galaxy. Barely-detectable magnetars ($L_{\rm X} = 3\times10^{33}$
\ergs\ [0.5--10.0 keV] and $A_{\rm rms}$$=$15\%) could have been
identified throughout $\approx$0.4\% of the spiral arms. The lack of
new examples implies that $<$540 exist in the Galaxy (90\%
confidence). Similar constraints are found by considering the
detectability of transient magnetars in outburst by current and past
X-ray missions. For assumed lifetimes of $10^{4}$ yr, we find that the
birth rate of magnetars could range between 0.003 and
0.06~yr$^{-1}$. Therefore, the birth rate of magnetars is at least
10\% of that for normal radio pulsars. The magnetar birth rate could
exceed that of radio pulsars, unless the lifetimes of transient
magnetars are $\ga$$10^5$~yr. Obtaining better constraints will
require wide-field X-ray or radio searches for transient X-ray pulsars
similar to XTE~J1810--197, AX~J1845.0--0250, CXOU~J164710.2--455216,
and 1E~1547.0-5408.  
\end{abstract} 

\section{Introduction}

Astronomers have only started to appreciate the diversity of the
properties of neutron stars that are produced when a massive star collapses
and explodes \citep{ptp06}. The list of different manifestations of
neutron stars now includes: radio pulsars that are powered by the
rotation of their $10^{8}-10^{13}$ G magnetic fields
\citep[e.g.,][]{lor06}; accreting X-ray pulsars \citep[e.g.,][]{bil97,wvdk98}
and thermonuclear bursters \citep[e.g.,][]{sb06} that are accreting
matter from binary companions; magnetars that are powered by the decay
of their $\ga$$10^{14}$ G fields \citep[which historically have been 
categorized as
either anomalous X-ray pulsars or soft gamma repeaters;
e.g.,][]{wt06}; intermittently-detectable ``Rotating RAdio
Transients'' \citep[RRATs;][]{mcl06}; isolated, cooling neutron stars
that shine primarily in soft X-rays \citep{wwn96,hab06}, and
central compact objects that are seen as point sources near the
centers of supernova remnants \citep[][]{cha01,sew03,pst04}.
 Understanding the properties of
these compact objects and their birth rates provides important constraints
on the late-time evolution of massive stars, and on the processes that occur 
during stellar collapse. The relationships 
among the different classes of compact object could reveal how their 
magnetic fields decay and their interiors cool.


In this paper, we present a search for magnetars. This search is
timely for two reasons. First, recent evidence suggests that magnetars are the
products of unusually massive progenitors. Three magnetars have been
found to be in clusters of massive, young stars \citep{fuc99,vrb00,eik04},
and the turn-off masses of two of these clusters imply that the
progenitors to the neutron stars were very massive, 30--40 \msun\
\citep{fig05,mun06}. A fourth magnetar has been associated with a
bubble of neutral hydrogen that was probably blown by the wind of a
$>$30 \msun\ progenitor \citep{gae05}.  This suggests that massive
stars may be more likely to produce magnetars, whereas ordinary
radio pulsars are generally presumed to be left by lower-mass, 8--20
\msun\ progenitors \citep[e.g.,][]{heg03}. Given that less massive 
stars are much more common
\citep[e.g.,][]{krou02}, if massive stars produce magnetars, one
would expect that their birth rates should be much lower than those of
radio pulsars \citep{gae05}.

Second, there is significant debate about how the strong 
magnetic fields that characterize magnetars are produced. The original 
hypothesis is that magnetars are born with millisecond periods, 
and that the strong fields are produced by a dynamo in the 
rapidly-rotating proto-neutron star \citep{td93,heg05}.  
However, observations of supernova remnants associated with magnetars
rule out the expected input of energy from neutron stars with initial
spin periods $\la$3 ms \citep{vk06}. At the same time, 
the discovery of a few OB stars with $10^3$ G surface fields
\citep{don02,don06a,don06b} has motivated the alternative hypothesis
that magnetar-strength fields are primordial, having been amplified only
by the collapse of the core \citep{fw06}. This second hypothesis is 
attractive because it makes a straightforward prediction, that the birth 
rate of magnetars should be equal to that of highly-magnetized OB stars. 
However, it cannot yet explain why some highly-magnetized neutron stars
are radio pulsars instead of magnetars \citep[e.g.,][]{pkg00}.
This discrepancy is one of the main points in favor of the 
$\alpha$--$\Omega$ dynamo process that would act in a rapidly-rotating 
proto-neutron star, because it could produce $\ga$$10^{15}$~G 
internal fields that would power the magnetars \citep{tlk02}.

Here, we report the results of our search for new Galactic magnetars in 
archival observations taken with the {\it Chandra X-ray Observatory} 
and the {\it XMM-Newton} Observatory. Previous searches for periodic
sources with these observatories have been limited to small fields, such
as the Small Magellanic Cloud \citep{mac03,edge04} and the central 20 pc
of the Galaxy \citep{mun03}. Our search incorporates observations through the 
entire Galactic plane. Although we find four new sources
with significant periodic signals, none are likely to be magnetars. 
Therefore, we use the survey to place limits on the number of active 
magnetars in the Galaxy, and discuss the implications for their birth 
rates. 

\begin{figure*}[htb]
\centerline{\psfig{file=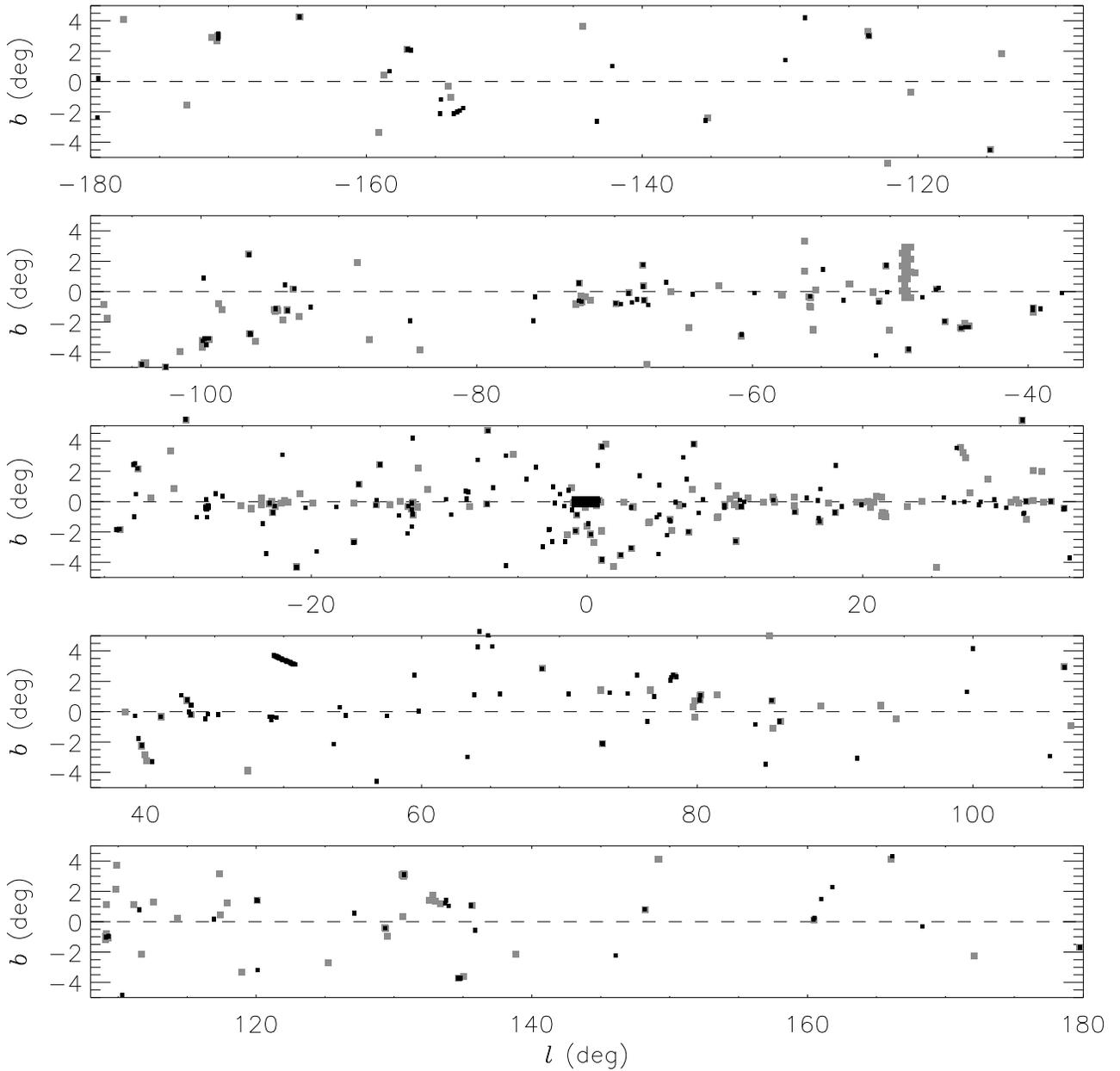,width=\linewidth}}
\caption{The locations of archival observations used for this survey.
Black squares are \chandra\ observations, schematically represented with
boxes that have an area equal to that of the ACIS-I. Grey rectangles 
are \xmm\ observations, schematically represented with boxes that
have an area equal to that of one EPIC MOS.}
\label{fig:coverage}
\end{figure*}

\section{Observations} 

We attempted to identify new magnetars by searching for sources with
periodic X-ray pulsations in archival \chandra\ and \xmm\ observations
of the Galactic plane. We included observations with a Galactic
latitude $|b|$$<$5\degree\ that were public as of January 2007.  For
both observatories, we required that some of the data were taken in an
imaging mode.  For \chandra, we further rejected observations taken
with the gratings in place. For \xmm, we rejected all observations
shorter than 10 ks, because they were often affected by background
flares throughout their entire duration. In total, we searched 506
\chandra\ observations, and 441 \xmm\ observations.  The coverage on
the sky is illustrated in Figure~\ref{fig:coverage}.  Within a
Galactic longitude of $|l|$$<$10\degree\ and a latitude
$|b|$$<$5\degree, which encloses about half of the mass of the Galaxy
(e.g., using the model in Launhardt, Zylka, \& Mezger 2002), these
observations cover about 6\% of the sky. Closer to the Galactic plane,
within $|l|$$<$10\degree\ and $|b|$$<$0.5\degree, the observations
cover 25\% of the sky.

The raw event lists and calibration data were downloaded
from the High Energy Astrophysics Science Archive Research 
Center.\footnote{{\tt http://heasarc.gsfc.nasa.gov/}}
We processed the event lists produced by each observatory in the standard
manner, in order to extract events for sources that we could search for
periodic variability. 

\subsection{Chandra Data Preparation}

The \chandra\ data were processed using CIAO
version 3.4. As we were only interested in the arrival times of events,
we did not apply the latest calibration. We started with the default
level 2 event files, and removed any time intervals during which flares 
from the particle background
caused the total event rate from the detectors to increase by more 
than 2$\sigma$ above the mean event rate. We then generated exposure maps
at a fiducial energy of 1.5 keV (i.e., the peak of the detector effective
area), and binned the event lists to produce images in the 0.5--8.0 keV band. 
These were used to search for point sources using the routine {\tt wavdetect} 
\citep{fre02}.  For computational efficiency, we searched
a series of three images using sequences of wavelet scales that
increased by a factor of $\sqrt{2}$: a central, un-binned image of
8.5\arcmin\ by 8.5\arcmin\ searched over scales of 1--4 pixels, an
image binned by a factor of two to cover 17\arcmin\ by 17\arcmin\
searched over scales of 1--8 pixels, and an image binned by a factor
of four to cover the entire field searched over scales of 1--16
pixels. The source lists from each image were combined for each
observation, favoring the positions derived from the images with the
highest spatial resolution for sources detected at multiple scales. We
did not attempt to discriminate real sources from detector artifacts,
such as the boundaries of the individual charge-coupled devices (CCDs), 
at this stage of the algorithm.

We then extracted events from each source, using a radius defined to
enclose 90\% of the point spread function (PSF) at 4.5 keV. In this
case, a relatively higher energy was used so that we would not exclude photons
from sources with hard spectra.  The radius was dependent on the
offset from the aim point in arcminutes ($\theta$), and was parameterized 
based on simulations with the CIAO tool {\tt mkpsf} as
\begin{equation}
r = 3.27 + 0.342\theta + 0.020\theta^2 + 0.019\theta^3,
\end{equation}
where $r$ is the radius in pixels (0\farcs492). 
We corrected the arrival times of the events to the Solar System barycenter
using the tool {\tt axbary}.

\subsection{XMM Data Preparation}

The \xmm\ event lists from the European Photon Imaging Camera (EPIC)
were analyzed using version 7.0 of the Science
Analysis Software,\footnote{{\tt
http://xmm.vilspa.esa.es/external/xmm\_user\_support
/documentation/sas\_wsg/USG/USG.html}} CIAO 3.4, and
the High Energy Astrophysics Software version 6.3.0.\footnote{{\tt
http://heasarc.gsfc.nasa.gov/docs/software/lheasoft/}}
We examined the lists from each 
active imaging detector separately, starting with the files from 
the archive.
For most observations, the same field was observed with three independent
cameras: one pn CCD array and two metal-oxide-silicon [MOS] CCD arrays.
We removed time intervals during which particle events caused the
event rate from the detector to flare more than two standard deviations
above the mean rate. This selection was generally successful at removing
particle flares, but in many cases the entire observation for a given 
camera was affected by flares, which rendered this automatic algorithm 
useless. We discarded data from individual cameras (usually the pn)
for observations too badly affected by flares.

We then created images of the 0.2--12 keV events, binned to 4\arcsec\
resolution. The standard data selection was applied to make the
images, in order to remove events near the edges of the detector chips
and bad pixels, and to reject events that were likely to be cosmic
rays (pattern $\ge$4 for the pn and $\ge$12 for the MOS). We then
generated matching exposure maps. We searched for point sources using
the routine {\tt ewavelet}, separately for each observation. At this
stage of the algorithm, we did not attempt to discriminate real
sources from detector artifacts, nor did we attempt to verify that
sources detected in one camera were present in the others.  We
extracted event lists from individual sources using the radius
produced by {\tt ewavelet}, which generally was $\approx$15\arcsec\ and
enclosed $\approx$50\% of the PSF.  
The arrival times of the
photons were corrected to the solar system barycenter using the tool
{\tt barycen}.

\subsection{Identifying Candidate Signals}

For sources with at least 100 total events from either observatory 
(including background), we computed Fourier periodograms to search for
periodic signals. The range of periods searched for both instruments was
designed to encompass those of known magnetars with X-ray pulsations, 
5--12 s \citep{wt06}. However, we note that after our analysis was completed, 
\citet{cam07} announced the discovery at radio wavelengths 
of 2~s pulsations from 1E~1547.0-5408.
Periods this short would generally only be identifiable in \xmm\ EPIC-pn
data with our search.

For the \chandra\ data, we used the Rayleigh statistic 
\citep[$Z_1^2$;][]{buc83}. We searched for signals with frequencies between 
1.5 times the Nyquist frequency (i.e., $1.5 \times 1/2t_{\rm bin}$, 
where $t_{\rm bin}$ is the interval at which the data were read out) and 
10\% of the inverse of the total time interval of the observation (i.e., 
$0.1/T_{\rm exp}$) to avoid red noise. We used a frequency step corresponding 
to the inverse of the total time interval ($1/T_{\rm exp}$).
For most observations,
the data were read out every $t_{\rm bin}$=3.1 s, so the highest frequencies
searched correspond to periods of 4.1 s. Observations typically lasted between
1.2 ks and 120 ks, so we could identify signals with  periods at the 
upper range of at least 120 s, and in some cases 12,000 s. 

For the \xmm\ data, the Rayleigh Statistic was computationally
inefficient to compute for pn observations of bright sources, so we
computed discrete fast Fourier transforms.  The data were padded so
that the number of points in the transform was a power of 2, so the
frequency resolution was generally finer than $1/T_{\rm exp}$. The
maximum frequency considered was the Nyquist frequency for the
data. The pn data was taken with a time resolution of at least 73.4
ms, providing sensitivity to periods as short as 0.15 s.  The MOS data
was taken with a time resolution of at least 2.4 s, and so our search
was sensitive to periods as short as 4.8 s. The lowest frequency
considered was 0.1/$T_{\rm exp}$.

The confidence with which we could exclude that any given signal was
produced by random noise depended upon the number of trial signals
examined, $N_{\rm trial}$.  The typical \chandra\ observation lasted
20 ks, and contained 3 sources with $>$100 counts, and had a time
resolution of 3.1 s, so that there were $N_{\rm trial}$$\approx$20,000
trial frequencies in the ensemble of periodograms from each
observation.  The typical \xmm\ observation also lasted 20 ks, and
contained $\approx$25 sources per camera with $>$100 counts. For each
MOS detector, with a time resolution of 2.4 s, the typical set of
periodograms contained $N_{\rm trial}$$\approx$200,000 trial periods
per observation. For the pn detector, with a time resolution of 73.4
ms, the typical set of periodograms contained $N_{\rm
trial}$$\approx$$1.8\times10^8$ trial periods. In total, we searched
$1.7\times10^{11}$ trial periods, the vast majority of which were from
the high-time-resolution data taken with the EPIC pn.

The powers produced by random noise in a periodogram in which $n$ measurements
have been averaged are distributed as a chi-squared function with $2n$ 
degrees of freedom. Following \citep{rem02}, we refer to the measured 
Fourier powers with $n$=1 as $P_{\rm meas}$, and normalize them so that 
the mean power produced by white noise is 1. The chance 
probability that noise would produce a signal larger than $P_{\rm meas}$ 
can be determined from an exponential distribution:
\begin{equation} 
{\rm prob} = 1 - (1-e^{-P_{\rm meas}})^{N_{\rm trial}} 
\approx N_{\rm trial} e^{-P_{\rm meas}},
\end{equation}
where the approximation is valid for $P_{\rm meas}$$\gg$1 \citep{rem02}.
Given the large number of trials for our entire search, a signal that
had a $<$0.1\% chance of resulting from noise must have a power 
$P_{\rm meas}$$>$32.8. Such a signal would be detected at a confidence level 
equivalent to 3$\sigma$ over the entire search, or 8$\sigma$ in a single 
trial.

However, a signal could also be considered significant if it was detected
with a lower power in multiple observations, as one might hope would occur
given that \xmm\ has three separate cameras that observe the same patch
of the sky. For instance, given two signals $P_{\rm meas,1}$ and 
$P_{\rm meas,2}$ at the 
same frequency, the chance probability that their sum exceeds some
value is a chi-squared distribution with 4 degrees of freedom:
\begin{equation}
{\rm prob} \approx N_{\rm trial} 
  (1 + P_{\rm meas,1} + P_{\rm meas,2}) e^{-(P_{\rm meas,1} + P_{\rm meas,2})}
\end{equation}
\citep[see][for the general form for summing an arbitrary number of signals]{rem02}. If we take $P_{\rm meas,1} = P_{\rm meas,2}$, for example, 
a signal has a $<$0.1\% of resulting from noise if it appeared with 
$P_{\rm meas,1}$$>$17 in both observations.
In principle, one could devise an algorithm that searched through all
of the periodograms from the same source, and sum the powers at each 
frequency to search for signals that repeat in the data. In practice,
however, the periodograms were not all computed with the same frequency 
resolution, which makes such an effort difficult. Moreover, when considering
observations separated in time, one also has to be concerned that some
candidate signals drifted in frequency, either because of the spin-down
of an isolated pulsar, or Doppler shifts for a pulsar in a binary 
\citep[see, e.g.,][for further discussion]{vau94}. 

Therefore, we have adopted a simplified approach in examining
candidate signals, by recording all signals with powers with less than
a 1\% chance of resulting from noise in a search of a {\it single
source}.  For \chandra\ ACIS, the threshold power is generally $P_{\rm
meas}$$>$17.  For the \xmm\ MOS, the threshold power is typically
$P_{\rm meas}$$>$19, whereas for the pn the power is $P_{\rm
meas}$$>$26. Any candidate signal was inspected to determine its
significance.

With these search criteria, our results up to this point were dominated
by signals that are non-periodic noise or detector artifacts. In both
\chandra\ and \xmm\ data, we detected low-frequency noise from astrophysical
flares in the count rates of individual sources such as
pre-main-sequence stars, and from background flares that our filtering
algorithm failed to remove. From \chandra\ ACIS, we detected signals
from sources that fell near chip boundaries, at the harmonics and beat
periods of the satellite dither (700s in the $x$-direction, 1000s in
the $y$). For the \xmm\ EPIC, particularly the pn, we found signals
with a range of periods that appeared to be related to hot columns and
chip boundaries, particularly in observations with high particle
background. We are not certain of the origin of these signals from the
EPIC.  The spurious signals introduced by the detector generally
shared the feature that they could be found in multiple sources at the
exact same frequency during an observation.  Therefore, we have
removed from consideration any signals that appeared in two or more
sources on the same detector in the same observation.  After removing
such signals, we found 358 sources with candidate signals in the 
\chandra\ observations, and 1380 sources (some of which are duplicates) 
with signals from \xmm.

These signals still turned out to be dominated by low-frequency noise
and detector artifacts, which could be quickly determined by visually
inspecting the power spectrum. Therefore, for the final step, we
scrutinized $\approx$1700 power spectra by eye to remove the remaining
examples that were clearly noisy, and to remove sources that appeared to
be detector artifacts. We defined a signal
as significant if it had a power $P_{\rm meas}$$>$32.8 in a single
observation or had a power larger than the single-observation
threshold in two or more observations.  We found a few sources with 
significant
periodic signals that we could not attribute to noise or detector
artifacts. We describe the previously-known and new sources separately
below.

\begin{deluxetable*}{lccccccc}[htb]
\tablecolumns{8}
\tablewidth{0pc}
\tablecaption{Candidate Periodic Signals\label{tab:signals}}
\tablehead{
\colhead{Source} & \colhead{ObsID} &  \colhead{Detector} & \colhead{Counts} & 
\colhead{$T_{\rm exp}$} & \colhead{Period} & \colhead{$P_{\rm meas}$} \\
\colhead{} & \colhead{} & \colhead{} & \colhead{} & \colhead{(ks)} & 
\colhead{(s)} & \colhead{} 
} 
\startdata
CXOU J174728.0--321445 & 4567 & ACIS-S & 2076 & 42.1 & 4910$\pm$20 & 199 \\ 
 & 4566 & ACIS-S & 1489 & 28.3 & 4790$\pm$50 & 81.9 \\ 
CXOU J182531.4--144036 & 5341 & ACIS-I & 548 & 18.0 & 780$\pm$3 & 36.4 \\ 
 & & & & & 5000$\pm$100 & 42.8 \\ 
\hline \\
XMMU J124429.7--630407 & 010948101 & pn & 491 & 49.0 & 475.0$\pm$0.6 & 21.3 \\
 & & MOS-1 & 187 & 52.0 & 474.6$\pm$0.8 & 14.4 \\
 & & MOS-2 & 162 & 52.0 & 474.5$\pm$1.0 & 3.5 \\
& 010948401 & pn & 403 & 39.3 & 475.1$\pm$0.5 & 23.5 \\ 
 & & MOS-1 & 145 & 41.7 & 476.7$\pm$1.7 & 11.1 \\ 
 & & MOS-2 & 141 & 41.1 & 473.9$\pm$1.2 & 11.8 \\ 
XMMU J185330.7--012815 & 0201500301 & pn & 20637 & 20.0 & 238.2$\pm$0.1 & 74.2 \\ 
 & & MOS-1 & 6781 & 20.1 & 238.4$\pm$0.2 & 39.1 \\ 
 & & MOS-2 & 6953 & 20.2 & 238.4$\pm$0.2 & 26.3   
\enddata
\tablecomments{The columns are as follows: the source names; the identifiers of
the observations in which the signals were found; the detector with which the
sources were observed; the total number of counts extracted for the source
(includes background); the exposure times of the observations; the periods,
which were computed by tracking the phases of the oscillations; and 
the power which the source was identified in the initial power spectrum.}
\end{deluxetable*}

\subsubsection{Known Sources}

Most of the signals were from previously-identified pulsars. These
confirmed that our algorithm worked as intended. In our \chandra\
observations, we identified the high-mass X-ray binary (HMXB)
4U~1145--619 \citep{whi78,rut05}, CXOU~J164710.2--455216
\citep{mun06}, and two sources toward the Galactic center
\citep[CXOC~J174532.3--290251 and GXOGC~J174532.7--290552;][]{mun03}.
In the \xmm\ observations, we identified the pulsars Geminga
\citep{hh92,jh05} and PSR~J1513--5908 \citep{sh82}, the HMXB Sct~X-1
\citep{koy91,kap07}, and the magnetars 1E~1048.1--5937
\citep{scs86,tie05}, 1RXS~J170849.0--400910 \citep{sug97,rea05},
SGR~1806-20 \citep{mur94,mer05}, SGR~1900+14 \citep{hur99,mer06},
XTE~J1810--197 in outburst \citep{ibr04,hg05}, and 1E~2259+586
\citep{fg81}.  The list above includes 7 of the 12 confirmed magnetars
in the Galaxy.

It is notable, however, that several magnetars were not detected in 
our search, despite being the targets of archival \chandra\ and \xmm\ 
observations. Here, we summarize the difficulties encountered identifying
several examples:
\medskip

\noindent
{\it 1E~2259+586} was not identified with \chandra\ because it saturated 
the detector during an imaging observation.

\noindent 
{\it 1E 1048.1--5937} and {\it 1RXS~J170849.0--400910} were not identified
with \chandra\ because they were only observed with the gratings in place.
These cases are not a serious concern, because such bright sources are
rare, and so almost never are found serendipitously in the fields of 
\chandra\ and \xmm. 

\noindent 
{\it SGR~1806--20} and {\it SGR~1900+14} were not identified by \chandra\
while in quiescence. Although their signals
were present in the data, their powers were below our search threshold.
All of the above sources were identified with \xmm.

\noindent
{\it 4U~0142+61} was not identified with either \chandra\ or 
\xmm. This is partly because the source had a small pulse fraction
(4\% rms), but also because of the detector modes with which
the source was observed. With \chandra\ the source either saturated 
the detector during imaging observations, or was observed with the gratings in 
place. With \xmm, only the MOS2 camera was active, and the magnetar 
only produced a signal above the single-observation
threshold ($P_{\rm meas}$$>$18) in one of the two observations. That signal
($P_{\rm meas}$=24.7) was below the threshold for our entire search, and
so can not be considered a detection as part of our blind search. 
However, had that source been observed
with the pn active, we would have identified it.

\noindent 
{\it 1E~1841--045} also  was not identified with \chandra\ or \xmm.
With a fractional rms amplitude of 13\% rms, it did not produce
a significant signal in the \chandra\ data. The \xmm\ observations of this 
source were too short ($<$10 ks) to be included in our search. A longer
\xmm\ observation would almost certainly have identified this source.

\noindent
{\it SGR~1627--41}, {\it AX~J1845.0--0258}, and {\it XTE~J1810--197} in 
quiescence were all too faint to produce detectable pulsations, even in 
searches targeted at their known or suspected spin periods 
\citep{got04, tam06, mer06b}.
These objects, and possibly the newly-identified magnetar 1E~1547.0-5408,
represent a class of magnetars from which pulsations could only be detected
intermittently, or perhaps not at all, in a search like ours.

\subsubsection{Newly-Identified Sources}

Four sources produced periodic signals that have not been previously
reported. In Table~\ref{tab:signals} we have listed basic information
about each source and signal. Figures~\ref{fig:cxcfft} and
\ref{fig:xmmfft} contain the Fourier power spectra in which the
candidate signals were discovered.  The power spectra provided initial
estimates of the pulse periods. We then refined the periods by (1)
computing pulse profiles from non-overlapping intervals between 1000
and 10,000 s long (depending upon the number of counts from the
source), (2) measuring their phases by fitting a sinusoid to the data,
and (3) modeling the differences between the assumed and measured
phases using first-order polynomials. The pulse profiles are displayed
in Figures~\ref{fig:cxcprof} and \ref{fig:xmmprof}. Below we remark on
some peculiarities to each source:

\begin{figure}
\centerline{\psfig{file=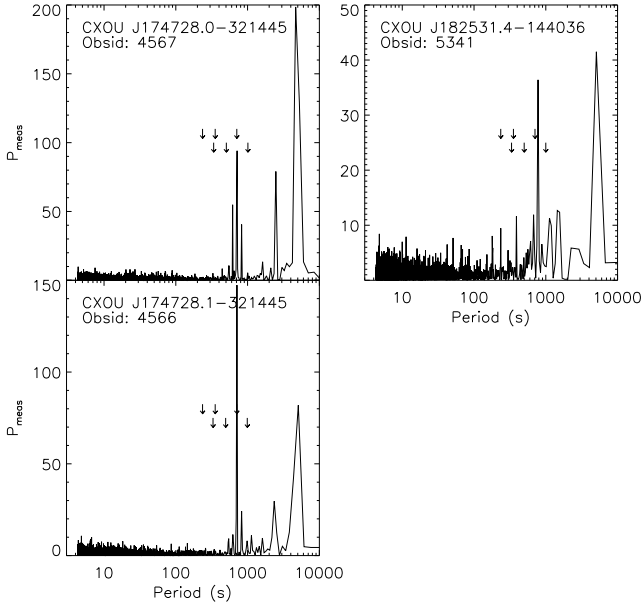,width=\linewidth}}
\caption{Power spectra from sources with periodic variability 
in archival \chandra\ observations. The left panels display power spectra
from the two observations of CXOU J174728.0--321445, and the right panel 
displays it from CXOU J182531.4--144036. The downward-pointing
arrows denote the fundamental and two harmonics of the $\approx$700 s
and $\approx$1000 s dither periods for the satellite. The $\approx$700 s
dither period, along with beat periods between that and the 5000 s signal,
are evident in the data from CXOU J174728.0--321445, because
the source lay at the edge of the detector.}
\label{fig:cxcfft}
\end{figure} 

\begin{figure}
\centerline{\psfig{file=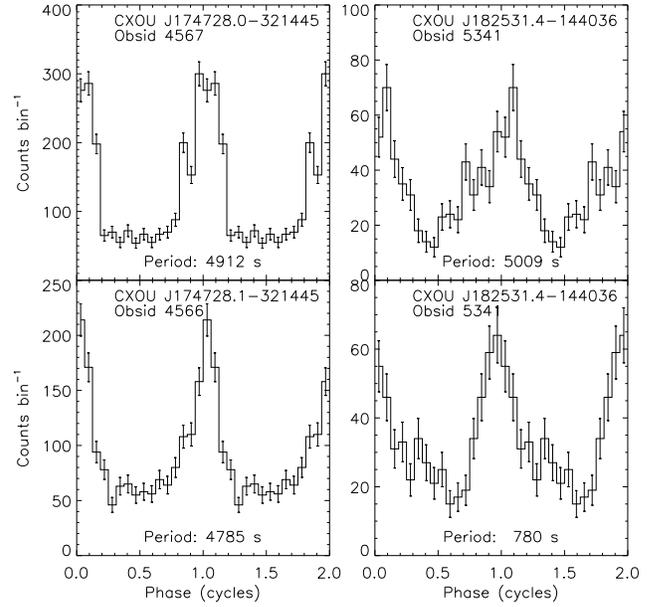,width=\linewidth}}
\caption{Profiles of the periodic signals detected in archival \chandra\ 
observations. Two cycles are repeated in each pane. 
The left panels display the profiles of the $\approx$5000~s signals 
from the two observations of CXOU J174728.0--321445. The right panels 
display the profiles of the 780 s (top panel) and 5000 s (bottom panel)
signals from CXOU J182531.4--144036. The period is printed at the bottom
of each panel.}
\label{fig:cxcprof}
\end{figure} 

\begin{figure*}[htb]
\centerline{\psfig{file=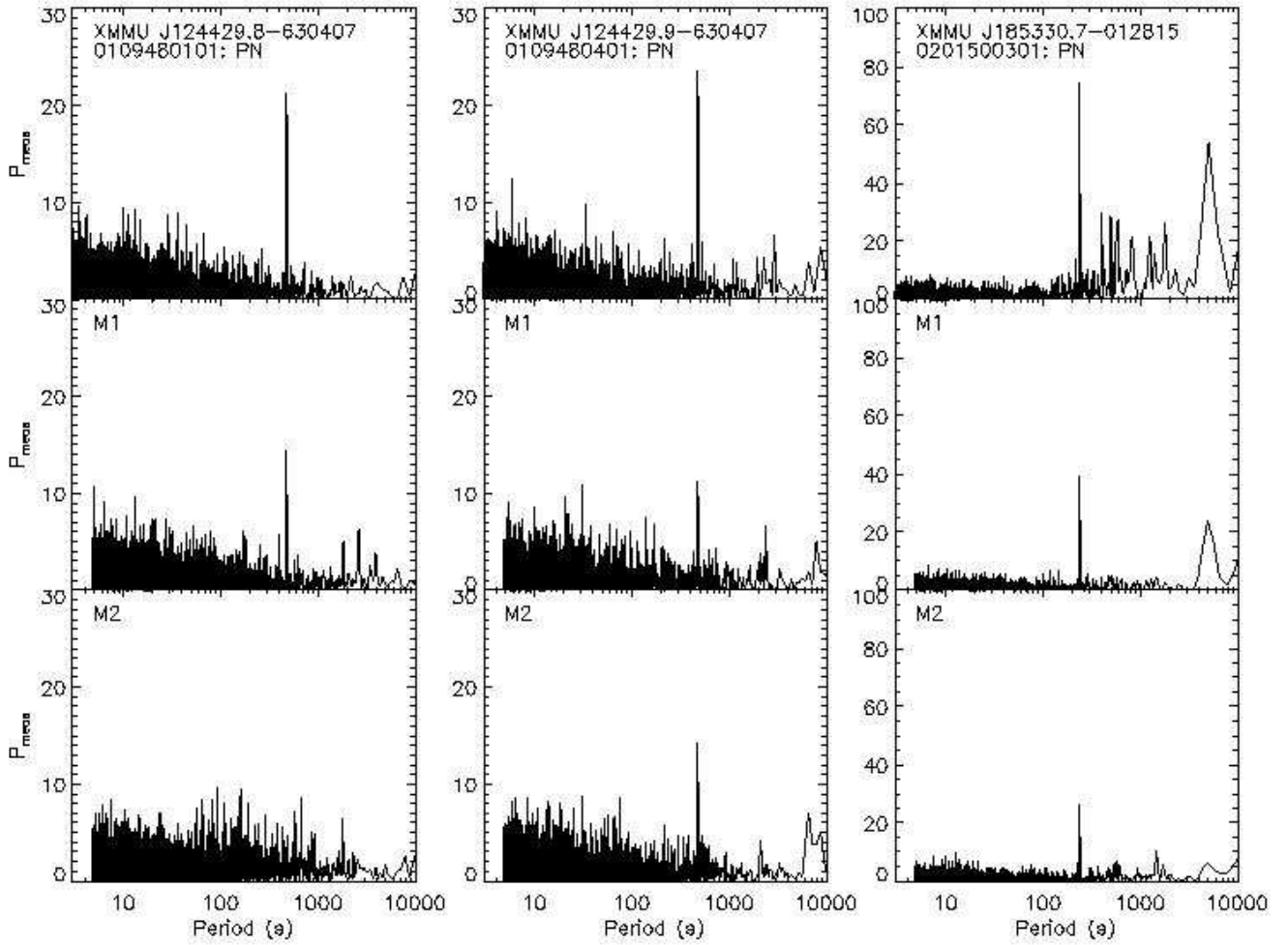,width=\linewidth}}
\caption{Power spectra from sources with periodic variability 
in archival \xmm\ observation. From left to right, we display power
spectra from the first and second observations of XMMU J124429.7--630407,
and from the one observation of XMMU J185330.7--012815. From top to bottom,
we display data from the EPIC-pn, MOS1, and MOS2.}
\label{fig:xmmfft}
\end{figure*} 

\begin{figure*}[htb]
\centerline{\psfig{file=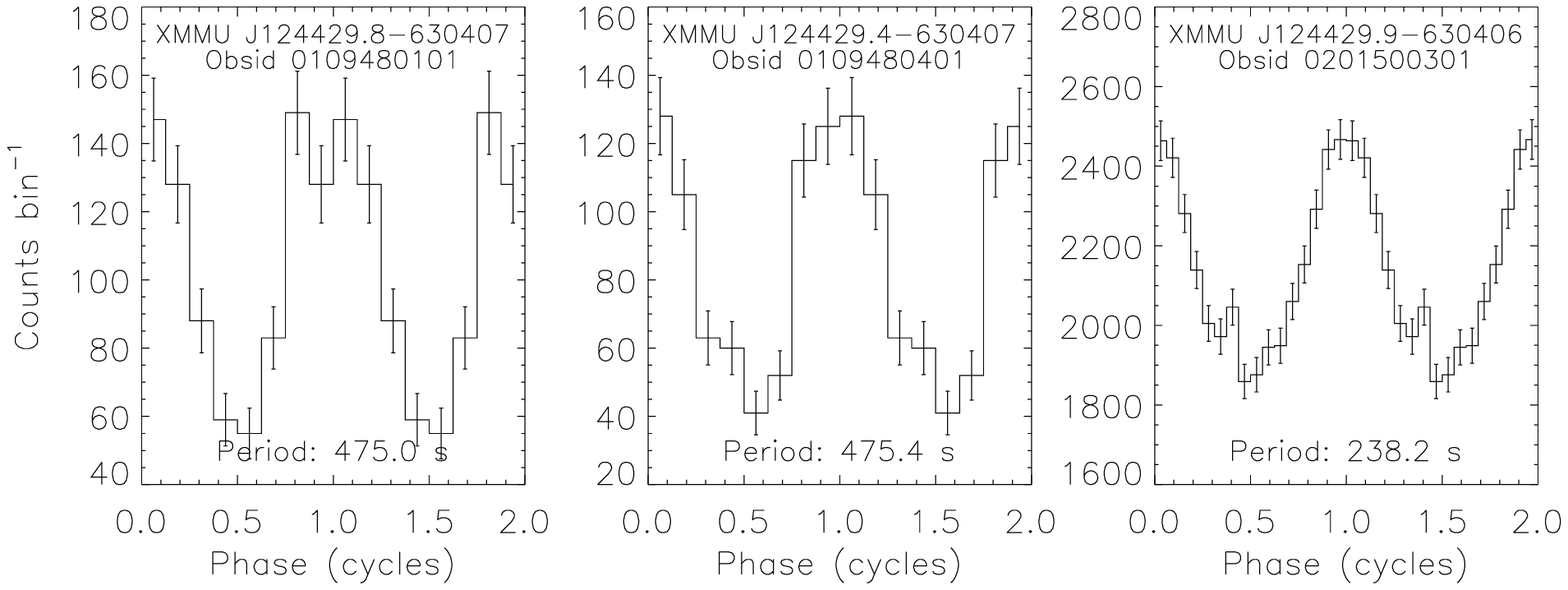,width=0.8\linewidth}}
\caption{Profiles of the periodic signals detected in archival \chandra\ 
observations. Two cycles are repeated in each panel, and data from 
all of the pn, MOS1, and MOS2 have been combined. 
From left to right, we display data from first and second observations 
of XMMU J124429.7--630407, and from the one observation of 
XMMU J185330.7--012815.}
\label{fig:xmmprof}
\end{figure*} 

\smallskip
\noindent
{\it CXOU J174728.0--321445}: The $\approx$5000 s period from this 
source is longer than 10\% of the exposure time, so it would not have 
been identified in our search were it not for the facts that the signal
has a strong harmonic near 2500 s. The source is detected in two 
observations. In both observations, it lies near the edge of a detector. 
The 707.4 s dither period is clearly detected in the data, but the 
$\approx$1000 s dither period (specifically, 1010.9 s in observation 4567 and 
989.9 s in observation 4566) is not seen because of the orientation of 
the detector. In addition,
signals are detected at the difference between the frequencies of the
dither and the 5000 s signal. The periods from the phase connection
analysis differ at the 2$\sigma$ level, so this signal may not be 
strictly coherent.

\noindent
{\it CXOU J182531.4--144036}: This source was identified for its 780~s
period. The signal is unrelated to the dither periods, which in this
observation were at 999.7~s and 707.6~s. Oddly, a $\approx$5000~s
period is also present in this source. We have confirmed that this
5000~s period does not appear in most sources observed with \chandra,
and the profile of the 5000~s signal is more sinusoidal in this source
than in CXOU J174728.0--321445.  However, only 4 cycles of the 5000~s
signal were covered by the observation, so this signal may be
low-frequency non-periodic noise that randomly produced a strong peak
in the power spectrum.

\noindent
{\it XMMU J124429.7--630407}: A 475~s period was detected from this source
in 5 of 6 trial power spectra from two different observations. The source is 
faint, and the non-detection with the MOS2 in observation 010948101 can
be ascribed to unfavorable noise. 

\noindent
{\it XMMU J185330.7--012815}: This source, also known as 
AX J1853.3-0128, exhibited a 238~s period in all three detectors in the one 
observation of it. The analysis of the phases of this source indicate
the signal is not coherent, as the phases vary by $\approx$0.1~cycles
with a time scale of $\approx$5000~s. The periodic signal from this source was 
identified independently by J. Halpern and E. Gotthelf, who also carried out
spectroscopy of its optical counterpart, and suggest that it is
a cataclysmic variable (private communication). 

\smallskip We note that it is curious that the period of the signal
from XMMU J124429.7--630407 is within 2$\sigma$ of being a factor of 2
longer than the period of the signal from XMMU J185330.7--012815.  We
have not been able to identify any systematic effect that might
explain both signals. In our search, we did not find similar
significant signals from any of the thousands of other sources that we
examined. The signals are unlikely to be detector artifacts, because
they are detected in the pn and both MOS cameras. Having failed to
identify any causes intrinsic to the spacecraft, on-board computers,
or data processing, we believe that they are
astrophysical, and that the factor of two difference between them is
simply a coincidence.

\section{Discussion}

The periodic signals that we have detected from previously-unidentified 
sources all have periods longer than 100 s. Although there is a chance
that some of these systems are neutron stars, we expect that most of 
them belong to the much-larger population of magnetically-accreting white 
dwarfs, either polars or intermediate polars 
\citep[e.g.,][]{nw89,sch02,rc03,mun03}.
Accreting white dwarfs typically have luminosities of 
$10^{30}-10^{32}$ \ergs, so that given their count rates, they could be
located as close as 50 pc (for XMMU J185330.7--012815) or as far
as 5 kpc (for XMMU J124429.7--630407). 
As members of the local Galactic neighborhood they will be 
distributed over the entire sky, not just within $|b|$$<$5\degree, 
our targeted survey of the Galactic
plane is not particularly efficient at finding cataclysmic variables. 
Therefore, we do not discuss them further. 

In the context of a search for magnetars, it is notable that
no new source was found to exhibit periodic variability with periods
in the range of known magnetars that are pulsed in X-rays, 5--12 s. 
To understand the lack of detections of obvious candidate neutron stars,
we need to compute the fraction of the Galaxy that was covered
by our survey. If the properties of magnetars as a population were 
bettern known, we would do so by assuming distributions for their luminosities
and pulse amplitudes, and carry out a maximum-likelihood or Monte-Carlo 
calculation to model the observed population \citep[e.g.,][]{fgk06, lor06}.
However, the intrinsic distributions for the luminosities and 
pulse fractions of magnetars are poorly constrainted. 
Some guidance can be obtained from models that explain the pulsations as
originating from single bright spots on their surfaces. \citet{ozel02}
find that the fractional amplitudes of pulsations are largest when the
hot spot is located on the equator, and viewed from the equator. However,
the amplitude drops by $\approx$50\% when the spot and viewer have a 
latitude $>$50\degree. Therefore, we roughly estimate that any given 
magnetar will be easily detectable over $\approx$65\% of the sky. 

Unfortuntely, as we describe below, the luminosities of magnetars are
highly variable, and cannot be predicted by first principles because
the mechanism causing the variability is not understood
\citep[e.g.,][]{wood04, mun07}. 
Therefore, in the following we only present some
representative cases. The total Galactic populations for our fiducial
examples are then calculated in two steps, first computing the depth
along the line-of-sight through the Galaxy to which our observations
were sensitive, and second estimating the fraction of the stellar mass
in the Galactic spiral arms that was enclosed by our observations.

\subsection{Depth of the Survey}

We compute the depth ($D$) through the Galaxy that each observation
was sensitive for any given luminosity ($L_{\rm X}$) and limiting
total number of counts ($C_{\rm lim}$) from: \begin{equation} C_{\rm
lim}/T_{\rm exp} = \xi_{\rm det}(N_{\rm H}[l,b,D]) \frac{L_{\rm
X}}{4\pi D^2}, \end{equation} where $\xi_{\rm det}(N_{\rm H}[l,b,D])$
is the conversion factor between flux and count rate for each
detector. This factor depends additionally upon the Galactic
absorption $N_{\rm H}$, which in turn is a function of the distance
$D$ to a source and its position in the Galactic plane, $l,b$.  We
computed $f(N_{\rm H})$ for both the \chandra\ ACIS-I and for the
\xmm\ EPIC pn behind the medium filter using the Portable, Interactive
Multi-Mission Simulator,\footnote{{\tt
http://heasarc.gsfc.nasa.gov/docs/software/tools/pimms.html}} assuming
several trial spectra and a range of $N_{\rm H}$. We
have neglected some other factors that do not significantly affect our
results, the choice of filters for \xmm\ (an $\approx$5\% effect on $\xi$), 
and hydrocarbon build-up on the ACIS (negligible for sources with $N_{\rm
H}$$\ga$$10^{22}$~cm$^{-2}$).

Vignetting as a function of offset from the aim point is a small
effect for \chandra, reducing the count rate from a source by
$\approx$20\% at an offset of 8\arcmin. We have neglected vignetting
in our \chandra\ observations.  However, it is a large effect for
\xmm, reducing the count rate by 50\% at an offset of
10\arcmin. Therefore, we have accounted for vignetting in our \xmm\
observations by reducing the flux-to-counts conversion $\xi$ by 33\%
from the on-axis value obtained from PIMMS, which is the mean
reduction over the inner 10\arcmin. We estimate that our simple
treatment of the vignetting introduces an uncertainty of $<$15\% on
the depth of the survey.

The spectra of
magnetars are generally described as the sum of a blackbody component
with a temperature of $kT$$\approx$0.6~keV and a soft power-law tail
with photon index $\Gamma$$\approx$3 \citep[e.g.,][]{wt06}.
The overall spectrum can be roughly approximated as a $\Gamma$=2
power law, so we take that as our fiducial spectrum.
Choosing instead a softer $\Gamma$=3 power law results in values of $D$ that 
are up to 30\% smaller (depending upon the absorption through the line
of sight; see below), while choosing a harder $kT$=0.6 keV blackbody 
increases $D$ by up to 20\%. 

We estimated the absorption as a function of 
distance along different lines of sight using models for
Galactic optical and infrared extinction. For most of the Galactic plane,
we linearly interpolated visual extinction values ($A_V$) from the model of 
\citet{dri03}, and converted them to $K$ band extinction values using
the relation $A_K/A_V = 0.11$ \citep{rl85,mat90}. However, within the
central 25\degree\ of the Galaxy the model extinction from
\citet{dri03} was significantly lower than that observed
\citep[e.g.,][]{lzm02}, so instead we interpolated values of the $K_s$
band extinction from the table in \citet{mar06}. We then converted the
$K$ band extinction (ignoring the 5\% difference in $A_K$ and
$A_{K_s}$) into a column density using $N_{\rm H} = 1.6\times10^{23}
A_K$ cm$^2$ \citep{rl85,ps95}.  Absorption through the Galactic plane
can reduce the flux observed from a source by up to a factor of 30 
compared to the value without absorption. We tested different values 
for $N_{\rm H}/A_K$ in the
range $(1.2-2.1)\times10^{22}$ cm$^2$
\citep[representing values from, e.g.,][]{glass99,td04}, and found that 
the uncertainty in $D$ introduced by our choice of absorption model is 
$\approx$15\%.

The depth of our survey depends strongly upon the assumed luminosity
and the fractional root-mean-squared (rms) amplitude of the pulsations
($A_{\rm rms}$) for which we are searching. The rms amplitude enters
consideration because it determines $C_{\rm lim}$: 
\begin{equation} 
A_{\rm rms} = 1.13 \left(\frac{2P_{\rm sig}}{C_{\rm lim}}\right)^{1/2}
{\rm sinc}^{-1}\left( \frac{\pi}{2}\frac{\nu}{\nu_{\rm Nyq}}\right),
\end{equation}
where $P_{\rm sig}$ is the intrinsic power of a detectable signal, the
factor of 1.13 is an average correction that accounts for the fact that
signals will often fall between independent Fourier bins, and the 
${\rm sinc}$ term takes into account the attenuation of the power of a signal
as its frequency $\nu$ approaches the Nyquist value $\nu_{\rm Nyq}$
\citep{vau94}. The intrinsic power $P_{\rm sig}$ detectable in a search
must be determined using the distribution of noise powers, as described in
\citet{vau94}. For a search threshold of $P_{\rm meas}$=32.8, the
expected value (50\% confidence) is $P_{\rm sig}$=31.8, 
and the power detectable in 90\% of trials is $P_{\rm sig}$=48.0. 
In the following
we will use the 90\% confidence level, $P_{\rm sig}$=48.0.
The value of $D$ computed for $P_{\rm sig}$=31.8 is $\approx$10--20\%
larger than the value for $P_{\rm sig}$=48.0. 

\begin{figure}
\centerline{\psfig{file=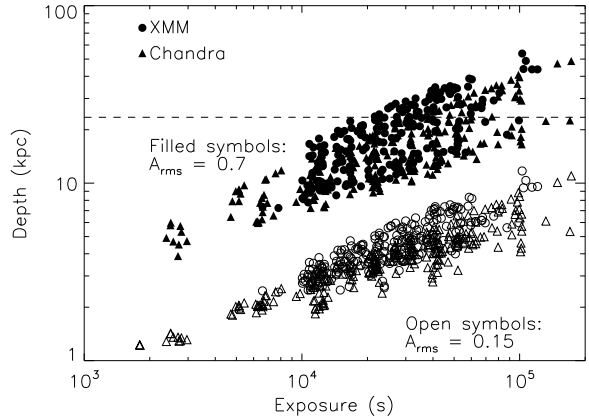,width=\linewidth}}
\caption{The depth through the Galaxy that each observation sampled, 
plotted as a function of its exposure time. Triangles illustrate
\chandra\ ACIS observations, and circles \xmm\ EPIC-pn observations.
The EPIC-pn has a larger effective area than ACIS, so the \xmm\ 
observations are systematically deeper for a given exposure. 
Filled symbols indicate the depth at which we are sensitive to 
an easily-detectable pulsar with $L_{\rm X} = 3\times10^{33}$ \ergs\ 
and $A_{\rm rms}$=70\%. Open symbols 
indicate the depth at which we are sensitive to 
a barely-detectable pulsar with $L_{\rm X} = 3\times10^{33}$ \ergs\ 
and $A_{\rm rms}$=15\%. The dashed line at a depth of 24 kpc 
demarks observations that proble the entire Galaxy; larger depths
are no longer meaningful because they extend beyond the distribution
of young stars.
}
\label{fig:depths}
\end{figure}

For our depth calculation, we use two extremes as examples. The first
is an easily-detectable magnetar. As our model for this, we use a
faint pulsar with a high pulse fraction, such as CXOU
J164710.2--455216 in Westerlund 1 \citep{mun06}, which had a
luminosity of $L_{\rm X}$=$3\times10^{33}$ \ergs\ (0.5--10 keV) and a
fully-modulated sinusoidal pulse profile with $A_{\rm
rms}$=0.71. Pulsations like this could be detected in 90\% of trials
with as few as $\approx$120 photons.  The depth probed by any given
observation scales as $L_{\rm X}^{1/2}$ and linearly with $A_{\rm
rms}$, so the easily-detectable example is equivalent to a bright
magnetar like SGR 1900+14, with $L_{\rm X}$=$10^{35}$ \ergs\
(0.5--10 keV)
and a sinusoidal pulse profile with $A_{\rm rms}$=0.12 \citep{hur99}. 
In the top
panel of Figure~\ref{fig:depths}, we plot the depth at which each
observation would have been sensitive as a function of exposure time
for our easily-detectable example. Only 10\% of the \chandra\
observations and 20\% of the \xmm\ observations probe the entire depth
of the Galactic plane to a distance of 24 kpc from Earth.

The other extreme is a barely-detectable pulsar, with a low luminosity
$L_{\rm X}$=$3\times10^{33}$ \ergs\ and a small pulse fraction $A_{\rm
rms}$=0.15. This example is chosen to have been just detectable in
observations comparable to those of XTE J1810--197 in quiescence, for
which $L_{\rm X}$=$2\times10^{33}$ \ergs\ (0.5--10 keV), and from
which pulsations were not detected to a limit of $A_{\rm rms}$$<$17\%
\citep{got04}.  A source with $A_{\rm rms}$=0.15 would require 2800
photons to be identified in 90\% of trials.  Consequently, a search
sensitive to our barely-detectable pulsar would cover a factor of
$\approx$5 less depth than a search for easily-detectable magnetars
like the one in Westerlund 1.  We display the depth of a survey for
a barely-detectable magnetar in the bottom panel of
Figure~\ref{fig:depths}. None of the archival observations would probe
the entire Galaxy when searching for our barely-detectable pulsar.

\subsection{Fraction of Young Stars Enclosed by the Survey}

In order to estimate the number of magnetars in the Galaxy and 
their birth rates, we need to know the fraction of the Galaxy
that we have meaningfully surveyed for magnetars. Magnetars 
have massive progenitors: neutron stars form from stars within 
inital masses $>$8\msun\ \citep{heg03}, and there is evidence 
that magnetars form from $>$30 \msun\ stars \citep{gae05,fig05,mun06}.
The lifetimes of magnetars are thought to be $\sim$$10^4$ yr 
\citep{kou99,ggv99},
so even if some magnetars receive kicks of $\sim$1000 km~s$^{-1}$, 
they will be found within $\sim$10 pc of their birth place. Most massive
stars are concentrated in the Galaxy's spiral arms, so we have calculated
the fraction of the stellar mass in the spiral arms that our observations 
have covered.

We use the model for the stellar mass in the spiral arms from 
\citet{wain92}.
The locations of the arms are defined as a logarithmic spiral:
\begin{equation}
\theta(R) = \alpha \log(R/R_{\rm min}) + \theta_{\rm min},
\end{equation}
where $R$ is the radial distance from the Galactic center,
$\theta$ is the azimuthal angle in radians (with $\theta$=0 along
the line connecting the Earth to the Galactic center, and 
0$\le$$\theta$$<$$\pi$ for $\sin l$$\le$0), $R_{\min}$ is the radial 
distance at which the arms start, $\theta_{\rm min}$ is the angle at
which the arms start, and $\alpha$ is the winding constant. We define
four main spiral arms, and one smaller, ``local arm.'' Each arm is assumed
to extend through an angle $\theta_{\rm ext}$, out to a maximum radial 
distance of $R_{\rm max}$=15 kpc. The values of
$\alpha$, $R_{\rm min}$, $\theta_{\rm min}$, and $\theta_{\rm ext}$ are
listed in Table~\ref{tab:spiral}, taken directly from \citet{wain92}.
The parameters assume the Earth is 8.5 kpc from the Galactic center.

\begin{deluxetable}{lcccc}
\tablecolumns{5}
\tablewidth{0pc}
\tablecaption{Spiral Arm Parameters\label{tab:spiral}}
\tablehead{
\colhead{Arm} & \colhead{$\alpha$} & \colhead{$R_{\rm min}$} & 
\colhead{$\theta_{\rm min}$} & \colhead{$\theta_{\rm ext}$} \\
\colhead{} & \colhead{(radians)} & \colhead{(kpc)} & \colhead{(radians)}
& \colhead{(radians)}
} 
\startdata
1 & 4.25 & 3.48 & 0.000 & 6.00 \\
1$^\prime$ & 4.25 & 3.48 & 3.141 & 6.00 \\
2 & 4.89 & 4.90 &  2.525 & 5.47 \\
2$^\prime$ & 4.89 & 4.90 & 5.666 & 5.47 \\
L & 4.57 & 8.10 & 5.847 & 0.55
\enddata
\end{deluxetable}

\begin{figure}
\centerline{\psfig{file=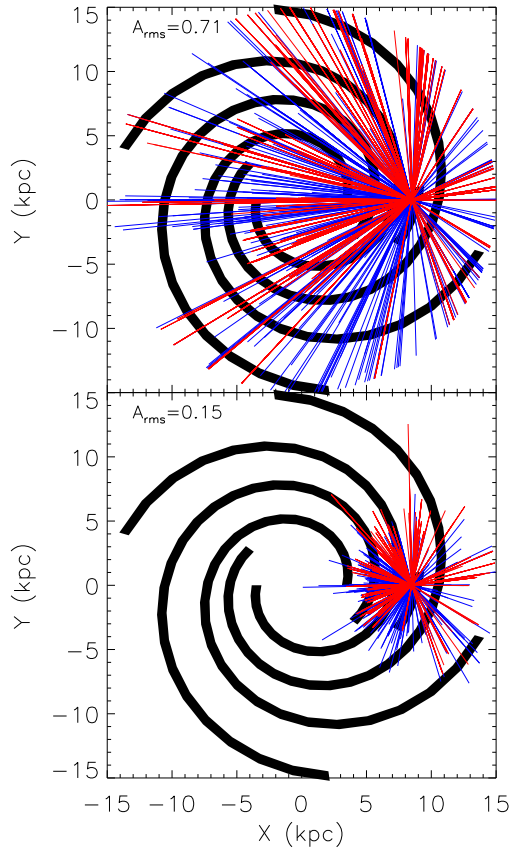,width=0.8\linewidth}}
\caption{The model for the Galactic spiral arms, viewed from above the
north Galactic Pole, overlaid by the depth of the observations in the survey.
The black lines are the model from \citet{wain92}. The blue lines represent
\chandra\ observations, and the red \xmm\ observations. The top
panel illustrate the depth to which we are sensitive to easily-detectable
pulsars, and the bottom to barely-detectable ones.}
\label{fig:spiral}
\end{figure}

The stellar density along the spiral arms is given by an exponential
distribution with radius $R$ and height above the plane $z$ \citep{wain92}:
\begin{equation}
\rho = \rho_0 \exp\left( - R/h - |z|/h_z \right),
\end{equation}
where $h$=3.5 kpc is the radial scale length, $h_z$=90 pc is 
the scale height of the youngest stars in the Galactic plane, and
$\rho_0$ is a normalization for which the exact value is unimportant, 
because we will divide by the total mass to obtain the fraction of the
arms encompassed by the survey. The density of the spiral arms is assumed
to be constant for any given $R$ within $\pm$375 pc of the center
of the arm, and zero outside that radial extent. This model for the
spiral arms is illustrated schematically in Figure~\ref{fig:spiral}.
In that figure, we also plot the depth through the Galaxy for which 
we would detect easily- and barely-detectable pulsars in 90\% of 
trials ($P_{\rm sig}$=48.0) for each of our archival observations.
The 50\% completeness level would incorporate about 20\% more of 
the Galaxy, but using it would add the complication that a larger 
fraction of magnetars with $L_{\rm X}$ and $A_{\rm rms}$ above our 
stated limits would be missed.

\begin{figure}
\centerline{\psfig{file=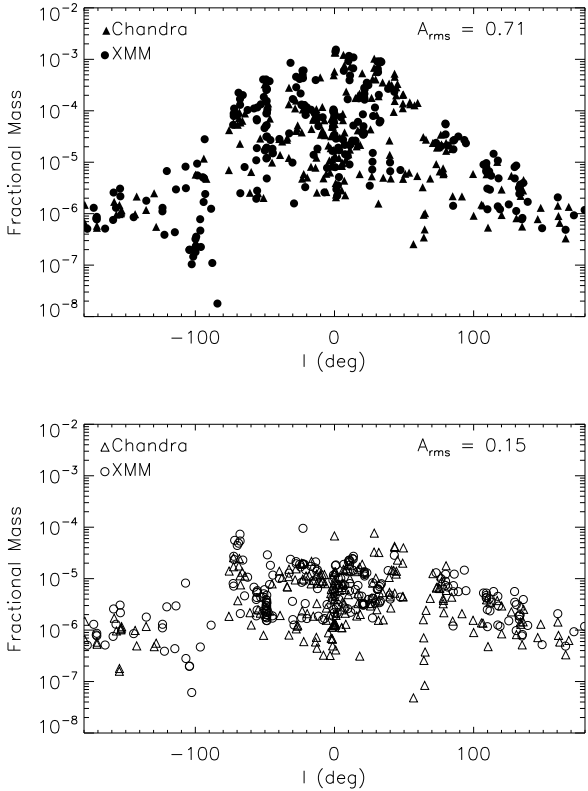,width=\linewidth}}
\caption{The fraction of the mass of the spiral arms surveyed by 
each observation, plotted as a function of Galactic longitude. 
Triangles illustrate
\chandra\ ACIS observations, and circles \xmm\ EPIC-pn observations.
The EPIC-pn has a larger effective area than ACIS, so the \xmm\ 
observations are systematically deeper for a given exposure. 
The top panel, with filled symbols, indicates the mass surveyed for which we
are sensitive to an easily-detectable pulsar, with 
$L_{\rm X} = 3\times10^{33}$ \ergs\ and 
$A_{\rm rms}$=70\%. The bottom panel, with
open symbols, indicates the mass surveyed for which we are sensitive to 
a barely-detectable pulsar with $L_{\rm X} = 3\times10^{33}$ \ergs\ and
$A_{\rm rms}$=15\%.
}
\label{fig:mass}
\end{figure}

For each observation, we integrated the mass enclosed along the line
of sight out to the depths defined in \S3.1. We assumed a field of
view of 17\arcmin$\times$17\arcmin\ for \chandra, which corresponds to
the full ACIS-I array. We assumed a field-of-view of
20\arcmin$\times$20\arcmin\ for \xmm, which corresponds to the region
where vignetting still provides $>$50\% of the on-axis count rate for
a source. The fractional mass of the spiral arms enclosed by each
observation is plotted as a function of Galactic longitude in
Figure~\ref{fig:mass}, for the cases of easily- and barely-detectable
pulsars.

To compute the total fractional mass enclosed by the survey, we
identified duplicate observations as those within 12\arcmin\ of each
other, and kept only the deepest of the duplicates.  Then, we summed
the fractional masses of the unique observations in
Figure~\ref{fig:mass}. We estimated uncertainties on the total
fractional masses by comparing the results of the calculations made
with different values for the fiducial spectrum and for the conversion
between infrared extinction and X-ray absorption (\S3.1).  We find
that our survey is 90\% complete ($P_{\rm sig}$=48.0) for
easily-detectable pulsars ($L_{\rm X}$=$3\times10^{33}$ \ergs, $A_{\rm
rms}$=71\%) for 5\% of the Galactic spiral arms under our standard
model. The \xmm\ and \chandra\ observations are about equally efficient, 
with \chandra\ surveying 2\% of the spiral arms, and \xmm\ 3\% of the spiral
arms.  Using
reasonable alternative spectra and absorption values described in
\S3.2, the total mass fraction surveyed can range between 3\% and
7\%. For barely-detectable pulsars ($L_{\rm X}$=$3\times10^{33}$
\ergs, $A_{\rm rms}$=15\%) our survey is complete for 0.4\% of the
spiral arms, with a range of 0.2--0.6\% if we choose different input
values. \xmm\ and \chandra\ each surveyed 0.2\% of the spiral arms on
their own.

We note that this model for the spiral arms does not include any young 
stars in the central 150 pc of the Galaxy, which has been surveyed by 
\citet{wgl02}
and Muno \etal\ (in prep). For the easily-detectable magnetar
case, our archival survey of this region reaches the Galactic center
(Figure~\ref{fig:spiral}).  This region contains $\sim$1\% of the
Galactic mass \citep{lzm02}, but the star formation rate is still
under debate. \citet{fig04} modeled the stellar population observed in
the infrared there, and concluded that the star formation rate is
$\sim$1\% of the Galactic value. However, it is possible that star
formation is skewed toward massive stars \citep{mor93}.  This is
suggested by indirect measurements of the Ly$\alpha$ flux in the
region, which could be 10\% of the total Galactic value
\citep{cl89,fig99}.  Depending upon the true star formation rate, the
central 2\degree$\times$1\degree of the Galaxy could contribute more than
half of the massive stars studied in this project.

\subsection{Constraints on the Population of Magnetars}

The main uncertainty in the birth rate of magnetars in the 
Galaxy is what the ``typical'' magnetar looks like. Most known 
examples are easily-identifiable in \chandra\ and \xmm\
observations, by which we mean that they are either: 
(1) luminous ($L_{\rm X}$$\sim$$10^{35}$ \ergs) 
with a modest pulse fraction ($A_{\rm rms}$$\ga$$13$\%) or (2) faint 
($L_{\rm X}$$\approx$$3\times10^{33}$ \ergs) with a large pulse fraction
($A_{\rm rms}$$\approx$$70$\%). Indeed 7 of 12 Galactic examples were 
identified blindly in our search (\S2.4). However,
pulsations are only intermittently-detectable from three transient magnetars 
--- SGR~1627--41 \citep{mer06b}, AX~J1845.0--0258 \citep{tam06}, 
and XTE~J1810--197 \citep{got04}. A fourth, 1E~1547.0--5408, has recently
been identified as a magnetar through its radio emission \citep{cam07}, 
but X-ray pulsations have not yet been confirmed \citep{gg07}.
The limits on their luminosities and pulse fraction serve as guides for 
our barely-detectable pulsars.  Therefore, there could
be a significant number of magnetars that can only be identified
in a blind search intermittently, when they produce outbursts.

For the purposes of this discussion, we will divide magnetars into 
three types: (1) the standard AXP-like sources, which are persistently
bright and can be detected with pointed observations at any time, (2)
the SGR-like magnetars, which can be identified by wide-field gamma-ray
burst monitors from anywhere in the galaxy whenever they produce outbursts
consisting of multiple soft gamma-ray bursts or occasional giant
gamma-ray flares, and (3) the transient AXPs, which can only be identified by 
pointed observations when they produce outbursts reaching a luminosity of 
$\ga$$10^{35}$ \ergs\ (0.5--10 keV). This is not to suggest that these groups
are mutually exclusive.  For instance, the standard AXP 1E~2259+586 
exhibits both short time scale bursts that appear similar to (although
much fainter than) SGR bursts, and variability in its mean intensity that
could be interpreted as a transient outburst with with time scales of months
\citep{wood04}. Moreover, in addition to its soft gamma-ray bursts, 
SGR 1627--41 also exhibits large (factor of 20) variations in 
its persistent X-ray luminosity between bursts \citep{wood99,hur00,mer06}, 
which could be 
interpreted as an outburst like those seen from transient AXPs
\citep{wood01, kou03, wood06}. Instead, we use these 
classifications to highlight the relative ease or difficulty with which 
different magnetars could be detected, and thereby to answer the question,
If there are many more magnetars in the Galaxy, what must they look like?

In the subsections below, we use the results of surveys most sensitive to 
each class of object. For standard AXPs, we use our own survey of
\chandra\ and \xmm\ data. For SGRs, the best constraints are provided 
by all-sky gamma-ray burst monitors. For transient AXPs, similar
results are found using our own archival survey, and past surveys with 
\rxte\ and \asca.

\subsubsection{The Number of Standard AXPs}

The standard AXPs are the easiest population to constrain,
because they are easily-detectable throughout the Galaxy. Our survey of 
archival \chandra\ and \xmm\ observations is 90\% complete for finding
standard AXPs for $\approx$5\% of the young stellar population in the 
Galaxy.

Before estimating a total number of standard AXPs, however, we need to 
establish how many known examples lie within our random survey of the Galaxy.
Only one source was identified in an observation in which it was not the
target: CXOU J164710.2--455216 in Westerlund 1 
\citep{mun06}. The other sources detected in our survey were observed 
as the targets of 
\chandra\ and \xmm\ observations because they were previously known, 
and some thought must be taken before they can be considered as part 
of a random sample. Three of the standard AXPS originally were identified
serendipitously during observations of other sources:
1E 1048.1--5937 during {\it Einstein} observations of the 
Carina nebula \citep{scs86}, 1E 2259+586 during 
{\it Einstein} observations of a supernova remnant  \citep{fg81}, 
and 1E 1841--045 during \asca\ observations of a 
supernova remnant \citep{vg97}. 
In the first case, the separation on the sky
between the magnetar and the central position of the original target 
were $\ga$0.5\degree, which makes it unlikely that the magnetar would 
have serendipitously fallen in the field of view of a \chandra\ or 
\xmm\ observation. In the latter two cases, the magnetars are close 
enough to the supernova remnants that one could have discovered them 
with \chandra\ or \xmm\ observations, if one makes the reasonable assumption
that the supernova remnant would be a sufficiently compelling target to 
have been observed. Therefore, we claim that 3 known magnetars 
lie in the random sample of the Galaxy that we surveyed. 

Knowing that there are 3 easily-detectable magnetars that can be found 
in our search of 5\% of the young stars in the Galaxy, we 
can determine the most likely total number of such objects in the Galaxy
using the binomial distribution. The probability that $n$ pulsars out of a 
total population $N$ would lie in a fraction $f$ of the Galaxy is:
\begin{equation}
{\rm prob}(n|f,N) = f^n (1-f)^{(N-n)} {{N!}\over{n!(N-n)!}}.
\label{eq:binomial}
\end{equation}
This can be inverted using Bayes' theorem to give the total number 
of magnetars $N$, given that $n$=3 magnetars are found with the fraction
$f$=0.05 of the Galaxy:
\begin{equation} 
{\rm prob}(N|f,n) = \frac{{\rm prob}(n|f,N)}{\sum_{N=n}^\infty {\rm prob}(n|f,N)},
\label{eq:bayes}
\end{equation}
The most likely value 
and 90\% confidence interval for 
the total number of easily-detectable magnetars is 
$59^{+92}_{-32}$.
Given that the number of known Galactic
magnetars is 12, there could be $47^{+92}_{-32}$ waiting to be discovered.

The Galactic rate of core-collapse supernovae is $\sim$0.01--0.04
yr$^{-1}$ \citep{cap93,vdbm94}, and the radio pulsar birth rate has
been recently estimated to be between $\approx$0.014~yr$^{-1}$
\citep{lor06} and $\approx$0.03~yr$^{-1}$ \citep{fgk06}. If magnetars
have typical lifetimes of $\sim$$10^{4}$~yr \citep{kou99,ggv99}, then
with somewhere between 27 and 160 in the Galaxy, the magnetar birth
rate would be between 0.003 and 0.016~yr$^{-1}$. The lower bound is
considerably smaller than the pulsar birth rate, as would be expected
if magnetars descend from $>$30\msun\ stars
\citep[e.g.,][]{gae05}. However, at the upper range, magnetars that
resemble the standard AXPs could have a birth rate nearly equal to
that of radio pulsars.

\begin{figure}
\centerline{\psfig{file=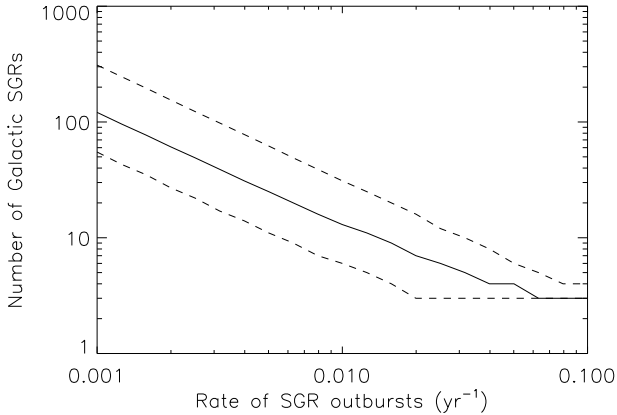,width=\linewidth}}
\caption{ The number of SGRs that could be present in the 
Galaxy, as a function of the rate at which year-long outbursts
(consisting of multiple, $\la$1~s soft gamma-ray bursts) recur.
The solid lines shows the most likely number, given that 3 Galactic SGRs
have been identified in the last 25 years, 
and the dashed lines show the 90\% confidence limit on that number.
}
\label{fig:numsgrs}
\end{figure}

\subsubsection{The Number of SGRs} 

An SGR can be identified if it produces multiple bright soft gamma-ray
bursts \citep[$\ga$$10^{40}$ \ergs, e.g.,][]{apt01}. Over the course of 
$\approx$25 years of monitoring, 3 Galactic SGRs have been confirmed 
\citep[and one in the LMC][]{wt06}.\footnote{In addition, one candidate
SGR has been proposed based on the identification of two bursts 
\citep{cline00}, and hundreds of individual SGR-like bursts have been 
identified all over the sky (see {\tt www.srl.berkeley.edu/iph3}).} 
Their bursts typically last $\sim$0.1~s
\citep{gog01}, so they are bright enough to be detected from anywhere 
in the Galaxy. Therefore, to estimate their total number, we only need 
to determine the rate at which SGRs produce detectable bursts. 

For any monitoring period of length $T$, if we define an ``outburst''
as containing multiple individual soft gamma-ray bursts, an SGR will 
be identified if it produces one or more outbursts. According to Poisson
statistics, if the outburst rate $r$ (where we take $r$ to have units such
that the expected number of outbursts in a time $T$ is $rT$), 
the chance that any given source is detected will be
\begin{equation}
{\rm prob}_{\rm det} = 1 - e^{-rT}.
\end{equation}
Given an ensemble of $n$ sources, the probability that $m$ sources will be detected (and $n-m$ not detected) is 
\begin{equation}
{\rm prob}(m|n,rT) = (1 - e^{-rT})^m (e^{-rT})^{(n-m)} {{n!}\over{m!(n-m)!}}.
\label{eq:poissprob}
\end{equation}
This is simply a binomial distribution, which can be inverted using Bayes'
theorem to obtain ${\rm prob}(n|m,rT)$ in the same way as 
Equation~\ref{eq:bayes}.

We can receive some guidance about the rate at which SGRs are active by noting
that, out of 25 years of monitoring, SGR~1806--20 has been active for 
$\approx$7 years, SGR~1900+14 for $\approx$4 years, SGR~0526--66 (in the LMC)
for $\approx$2 years, and SGR 1627--41 for $\approx$2 months 
\citep{apt01,wt06}.
The bursts tend to be associated with periods of activity lasting months to 
years. It is clear that the activity level is variable from source to source, 
so we consider a range of rates. 

Given $m$=3 and $T$=25 years, in Figure~\ref{fig:numsgrs} we plot 
for a range of $r$ the most-likely value and 90\% confidence limit 
for $n$. If on average one outburst is expected every 10 years, which is 
near the mean of the sample that has already been identified, then with 90\% 
confidence we should have already detected all of the Galactic SGRs. 
However, if an SGR only becomes active once per 100 years, there could
be $13^{+18}_{-7}$ in total in the Galaxy. Our best estimate is therefore 
that the number of Galactic SGRs is much smaller than that of standard
AXPs, unless SGRs are active less frequently than once per several hundred
years.

\begin{figure}
\centerline{\psfig{file=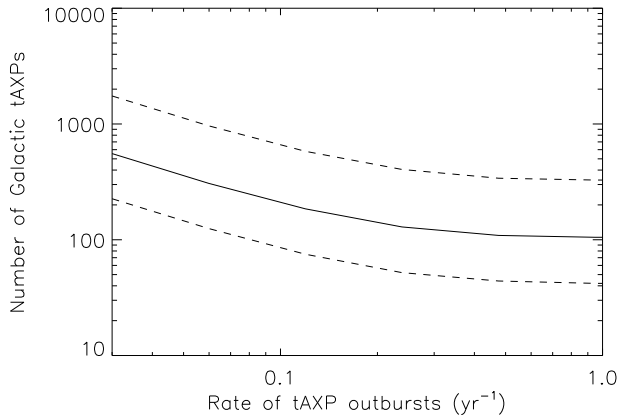,width=\linewidth}}
\caption{The number of transient AXPs that could be present in the
Galaxy, as a function of the rate at which year-long outbursts (in
which the X-ray luminosity is at least $10^{35}$ \ergs) recur.  The
number of Galactic examples is calculated based on observations with
\asca, \rxte, \chandra, and \xmm, which surveyed $\approx$2\% of the
Galaxy each year for 11 years, and discovered two transient
AXPs.  The solid lines shows the most likely number, and the dashed
lines show the 90\% confidence limit.  }
\label{fig:numtaxps}
\end{figure}

\subsubsection{The Number of Transient AXPs}

Transient AXPs would be the most difficult to identify, and could
comprise the majority of the magnetars. We define an AXP as transient
if it has a luminosity of $\la$$10^{33}$ \ergs\ (0.5--10 keV) for most
of its current life, but increases in luminosity to $\sim$$10^{35}$ \ergs\
for durations of on order a year \citep[e.g.,][]{got04,tam06}. We define 
the ``outburst'' as the year-long period when the AXP is luminous. 
Three transient AXPs are known (XTE~J1810--197,  
CXOU~J164710.2--455216, and 1E~1547.0--5408), along with one
candidate (AX~J1845.0--0258). 

Our archival \chandra\ and \xmm\ survey provides the most sensitive
search for transient AXPs in quiescence. Nonetheless, when looking for
a pulsar with $L_{\rm X}$=$3\times10^{33}$ \ergs\ (0.5--10 keV) and 
$A_{\rm rms}$=15\%, our survey is only 90\% complete for $\approx$0.4\%
of the Galactic young stellar population. The simple calculation
(Eq.~\ref{eq:bayes}) assuming no new barely-detectable magnetars in 
$f$=0.005 of the Galaxy,\footnote{CXOU~J164710.2--455216 is transient,
but was already counter among the easily-detectable examples, because it
has a fully-modulated pulse profile that is not characteristic of the
other transient AXPs.} implies that with 90\% confidence there could be 
up to 540 waiting to be discovered. Unfortunately, this estimate
is not a strict upper bound, because quiescent AXPs could be less 
luminous or have lower pulsed fractions than we have assumed.

An AXP in outburst could be detected over a larger fraction of the Galaxy, 
but it still requires an observation with a pointed X-ray observatory. 
To constrain their numbers based on AXPs in outburst, we must account 
for both the fraction of the young stars in the Galaxy that 
can be surveyed by a given observatory (Eq.~\ref{eq:binomial}), and
the chance that AXPs in the region will produce an outburst 
(Eq~\ref{eq:poissprob}). We assume that there is a population of
$N$ magnetars in the Galaxy with an outburst rate $r$, and that
we searched for them with a survey of duration $T$ that covered a fraction
$f$ of the magnetar birth places. Then, the probability that $m$
magnetars will be found is the joint probability that $n$ out of 
$N$ magnetars will lie in the survey region, and that $m$ out of $n$
will produce outbursts: 
\begin{eqnarray}
\nonumber {\rm prob}(m|rT,f,N) & = & \\
\nonumber  \sum_{n=m}^{N} & & (1 - e^{-rT})^m (e^{-rT})^{(n-m)} \frac{n!}{m!(n-m)!} \\
& & \times f^n (1-f)^{(N-n)} \frac{N!}{n!(N-n)!}.
\end{eqnarray}
This can be inverted using Bayes' theorem, in order to derive the probability 
that there are $N$ magnetars given that $m$ are found with outbursts
at a rate $r$ in a fraction $f$ of the Galaxy, ${\rm prob}(N|m,rT,f)$, as
in Equation~\ref{eq:bayes}. Here, because we are dealing with transient 
sources, $f$ represents the fraction of the young Galactic population searched
per year, which is the duration of a typical outburst.

We can estimate the rate of outbursts from transient AXPs from the 
four known examples (here we assume that the candidate is a genuine 
example). AX~J1845.0--0258 and XTE J1810--197 have each exhibited only
one outburst each, in 1993 and 2002 respectively, which in both cases led 
to their discoveries \citep{tor98, ibr04, got04, tam06}. In contrast,
CXOU~J164710.2--455216 entered into outburst only a year after its 
discovery \citep{mun06,mun07,isr07}, and 1E~1547.0--5408 has brightened to
$\sim$$10^{35}$ \ergs\ (0.5--10 keV) on at least two occasions in the last 
10 years \citep{gg07,cam07}. Taken together, these results suggest that the 
mean recurrence time is about once every 8 years. As we did for SGRs, we 
consider below a range of rates around this value.

To compute the fraction of the young Galactic population
that has been searched for transient AXPs, we need to take into account how
the known examples were discovered.
Two transient AXPs were discovered serendipitously: XTE J1810--197 in 
an \rxte\ observation of SGR 1806-20 \citep{ibr04}, and AX J1845.0--0250 in an 
\asca\ observation of Kes 75 \citep{tor98}. 
We consider these to be useful examples for estimating the total 
Galactic population. The other two probably would not have been idenitified
as transient AXPs were it not for unusual circumstances. The outburst 
from CXOU~J164710.2--455216 was identified because it produced a 
hard X-ray burst that was detected by {\it Swift}, but that burst
probably would have been ignored if the source had not been previously 
identified in quiescence \citep{isr07}. We have taken into account 
objects like CXOU~J164710.2--455216 to the Galactic population in
\S3.3.1, so we do not include it here. 1E 1547.0--5408 was positively 
dentified as a magnetar through radio observations of a variable, 
point-like X-ray source in a supernova remnant that was studied 
deliberately \citep{gg07,cam07}. X-ray pulsations have not yet been 
confirmed from this source, so we do not consider it here.

Based on the above considerations, we now calculate the fraction of
young stars that \rxte, \asca, \chandra, and \xmm\ surveyed 
per year during their
lifetimes, in the same manner as in \S3.1--3.2. We assume that the
target magnetar would have $L_{\rm X}$=$10^{35}$ \ergs\ (0.5--10 keV),
and a net number of counts determined by the observations in which
XTE~J1810--197 and AX~J1845.0--0258 were identified. For \rxte\,
XTE~J1810--197 was detected with $\approx$6 counts s$^{-1}$ in a 2.6
ks observation with the Proportional Counter Array
\citep[PCA;][]{ibr04}, so we assume that a pulsar would require
$1.6\times10^4$ total counts to be identifiable. We take the PCA field-of-view
to be 50\arcmin$\times$50\arcmin, and use a count-to-flux conversion that is
33\% of the on-axis value. We find that the
\rxte\ PCA has surveyed $\approx$2\% of the Galaxy down to this
sensitivity during the last 11 years, and $\approx$0.7\% in any given
year.  For \asca\, AX~J1845.0--0258 was detected with $\approx$1000 net counts
in each Gas Imaging Spectrometer \citep{tor98}. Vignetting drastically 
reduces the count rates from sources off-axis, so we assume a field-of-view
of only 20\arcmin$\times$20\arcmin, and use a flux-to-count conversion that
is 33\% smaller than the on-axis value.  We find that \asca\
observations were sensitive to bright magnetars over $\approx$3\% of the 
Galaxy during its 7 year lifetime, and $\approx$1\% of the Galaxy per year. 
Finally, we find that \chandra\ surveyed $\approx$0.5\% of the 
Galaxy per year over the last 7 years, and that \xmm\ surveyed 
$\approx$1\% of the Galaxy per year over the last 6 years. Taken together,
and accounting for overlaps, we find that over the last 11 years (since 
the launch of \rxte), $\approx$2\% of the Galaxy has been surveyed by 
a pointed X-ray observatory each year. 

In Figure~\ref{fig:numtaxps}, we plot the constraints on the total number
of transient AXPs, given that two were identified during observations
of $f$=0.02 of the young Galactic population each year for $T$=11 yr. 
Transient
AXPs are probably the largest undiscovered population of magnetars. 
For instance, if the average transient AXP produces outbursts at a rate of 
one per 10 years, then there could be $190^{+390}_{-110}$ of them in 
the Galaxy. The upper bound is comparable to the number of quiescent 
magnetars that we
estimate could be present based on archival \chandra\ and \xmm\ observations.
For a lifetime of $10^{4}$ yr \citep{kou99,ggv99}, this implies a brigh rate
of 0.008--0.06 yr$^{-1}$. The lower bound is $\approx$50\% of the radio 
pulsar birth rate, whereas the upper bound exeeds the higher estimate of 
the pulsar birth rate from \citet{fgk06}. The higher rate would be large
compared to the estimated Galactic supernova rate. However, it is possible
that transient magnetars have significantly longer lifetimes than have
been estimated based on the energy budgets of the persistent examples
\citep[e.g.,][]{kou99,ggv99}, in which case the birth rates would 
be correspondingly lower.

\section{Conclusions}

We have serached archival \chandra\ and \xmm\ observations for X-ray
pulsars, in order to constrain the Galactic population of
magnetars. Although we found four objects with periodic variability on
time scales of 200--5000~s (Figs.~\ref{fig:cxcfft}--\ref{fig:xmmprof}
and Table~\ref{tab:signals}), we found no sources that were obviously
magnetars. The archival observations that we used covered a moderate
fraction of the young stellar populations from which magnetars
descend. Our search was sensitive to the standard, easily-detectable
AXP-like magnetar ($L_{\rm X}$=$10^{35}$ \ergs\ [0.5--10 keV]
and a sinusoidal pulse profile with $A_{\rm rms}$=0.12, or
equivalently $L_{\rm X}$=$3\times10^{33}$ \ergs\ and $A_{\rm rms}$=0.7)
with 90\% completeness for $\approx$5\% of the mass in the Galactic 
spiral arms. Searching for transient AXPs in quiescence 
($L_{\rm X}$=$3\times10^{33}$ \ergs\ and $A_{\rm rms}$=0.15), however, 
we were only sensitive over 0.4\% of the Galactic spiral arms.

Based on the number of known Galactic magnetars, we then placed constraints
on the total population. Our archival search placed strict constraints
on the number of standard, persistent AXPs, which number 
$59^{+92}_{-32}$ with 90\% confidence. We also placed constraints on the 
number of SGR-like magnetars as a function of the rate at which they 
produce outbursts (Fig.~\ref{fig:numsgrs}), and find that for a likely 
recurrence rate of 0.01 $yr^{-1}$, there would be $13^{+18}_{-7}$ in 
the Galaxy. However, these populations may be dwarfed by the number of 
transient AXPs similar to XTE~J1810--197 \citep[see also][]{ibr04}. 
No new examples 
were detected in archival \chandra\ and \xmm\ observations, so up to 540
quiescent AXPs could be present in the Galaxy. 
We also considered the fact that two transient AXPs have been discovered
in outburst by \asca\ and \rxte. The fact that \asca, \rxte, \chandra, and
\xmm\ together have surveyed 2\% of
the Galaxy each year for the last 11 years allows us to constrain the 
number of transients as a function of their recurrence time 
(Fig.~\ref{fig:numtaxps}). For a recurrence time of once every 10 
years, this calculation suggests there are between 80 and
580 transient AXPs, with 90\% confidence. Assuming a lifetime of 
$10^4$ yr for persistent magnetars \citep{kou99,ggv99}, their birth rate 
is at least 10\% of that of radio pulsars, and could equal the radio pulsar
birth rate \citep[e.g.,][]{lor06,fgk06}. The birth rate of transient 
magnetars could exceed those of radio pulsars, unless their lifetimes are 
significantly longer than $10^4$ years. 

The main uncertainty in the number of Galactic magnetars is introduced
by the unknown activity level of transient examples. Further monitoring
of known transient AXPs would constrain their duty cycles and reduce this
uncertainty, so long as the known examples can be taken as representative
of all magnetars. Ultimately, though, a wide-field monitoring campaign is 
needed to identify more transient AXPs in outburst. This can be accomplished 
in principle in the X-ray band, although the sensitivity of current and
planned wide-field monitors 
\citep[$\approx$$2\times10^{-11}$ \ergcms, 0.5--10 keV;][]{lev96,rem00,bm06}
is only sufficient to barely detect a $10^{35}$ \ergs\ magnetar to a distance 
of 4 kpc from Earth. On the other hand, the detections of XTE J1810--197 
\citep{cam06} and 1E~1547.0--5408 \citep{cam07} in the radio with 
characteristic intensities of $\ga$100 mJy kpc$^{2}$ suggests that future
radio surveys with the Low-Frequency Array, the Mileura Wide-field Array,
or the Square Kilometer Array \citep[e.g.,][]{clm04} 
could provide the best constraints on
the numbers of magnetars in the Galaxy.

\acknowledgments
We thank E. Gotthelf for sharing the results of his observations of 
AX~J1853.3--0128. This work made use of data obtained from the
High Energy Astrophysics Science Archive Research Center (HEASARC), 
provided by NASA's Goddard Space Flight Center.

\end{document}